
\input epsf.tex
\def\pmb#1{\setbox0=\hbox{#1}%
  \hbox{\kern-.025em\copy0\kern-\wd0
  \kern.05em\copy0\kern-\wd0
  \kern-0.025em\raise.0433em\box0} }

\def \mapright#1{\smash{\mathop{\longrightarrow}\limits^{#1}}}
\def\half {{1\over 2}}
\def\eps{\tens{\hbox{\mm \char'42}}_{\hskip -5pt\infty}}
\catcode`@=11
\def\leftrightarrowfill{$\m@th\mathord\leftarrow \mkern-6mu
  \cleaders\hbox{$\mkern-2mu \mathord- \mkern-2mu$}\hfill
  \mkern-6mu \mathord\rightarrow$}
\def\overleftrightarrow#1{\vbox{\ialign{##\crcr
     \leftrightarrowfill\crcr\noalign{\kern-1pt\nointerlineskip}
     $\hfil\displaystyle{#1}\hfil$\crcr}}}
\catcode`@=12
\def\tens{\overleftrightarrow}

\def\approxge{\hbox {\hfil\raise .4ex\hbox{$>$}\kern-.75 em
\lower .7ex\hbox{$\sim$}\hfil}}
\def\approxle{\hbox {\hfil\raise .4ex\hbox{$<$}\kern-.75 em
\lower .7ex\hbox{$\sim$}\hfil}}

\def \abstract#1 {\vskip 0.5truecm\sepline\vskip 0.5truecm
$$\vbox{\hsize=15truecm\noindent #1}$$}
\def \SISSA#1#2 {\vfil\vfil\centerline{Ref. S.I.S.S.A. #1 CM (#2)}}
\def \PACS#1 {\vfil\line{\hfil\hbox to 15truecm{PACS numbers: #1 \hfil}\hfil}}

\def \hfigure
     #1#2#3       {\midinsert \vskip #2 truecm $$\vbox{\hsize=14.5truecm
             \seven\baselineskip=10pt\noindent {\bcp \noindent Figure  #1}.
                   #3 } $$ \vskip -20pt \endinsert }

\def \hfiglin
     #1#2#3       {\midinsert \vskip #2 truecm $$\vbox{\hsize=14.5truecm
              \seven\baselineskip=10pt\noindent {\bcp \hfil\noindent
                   Figure  #1}. #3 \hfil} $$ \vskip -20pt \endinsert }

\def \vfigure
     #1#2#3#4     {\dimen0=\hsize \advance\dimen0 by -#3truecm
                   \midinsert \vbox to #2truecm{ \seven
                   \parshape=1 #3truecm \dimen0 \baselineskip=10pt \vfill
                   \noindent{\bcp Figure #1} \pretolerance=6500#4 \vfill }
                   \endinsert }

\def\hfigureps#1#2#3#4{
\midinsert
      \vskip #2 truecm
      \special{#4}
      $$
         \vbox{
            \hsize=14.5truecm \seven\baselineskip=10pt\noindent
            {\bcp \noindent Figure  #1}. #3
         }
      $$
      \vskip -10pt
\endinsert
}
%
\def \ref
     #1#2         {\smallskip \item{[#1]}#2}
\def \sepline     {\medskip\centerline{\vbox{\hrule width5truecm}} \medskip}
\def \tabrule     {\noalign{\vskip 5truept \hrule\vskip 5truept} }
\def \tabrul2     {\noalign{\vskip 5truept \hrule \vskip 2truept \hrule
                   \vskip 5truept} }


\footline={\ifnum\pageno>0 \tenrm \hss \folio \hss \fi }

\def\today
 {\count10=\year\advance\count10 by -1900 \number\day--\ifcase
  \month \or Jan\or Feb\or Mar\or Apr\or May\or Jun\or
             Jul\or Aug\or Sep\or Oct\or Nov\or Dec\fi--\number\count10}

\def\hour{\count10=\time\count11=\count10
\divide\count10 by 60 \count12=\count10
\multiply\count12 by 60 \advance\count11 by -\count12\count12=0
\number\count10 :\ifnum\count11 < 10 \number\count12\fi\number\count11}

\def\draft{
   \baselineskip=20pt
   \def\makeheadline{\vbox to 10pt{\vskip-22.5pt
   \line{\vbox to 8.5pt{}\the\headline}\vss}\nointerlineskip}
   \headline={\hfill \seven {\bcp Draft version}: today is \today\ at \hour
              \hfill}
          }

%
%

%
\catcode`@=11
%
%
\def\b@lank{ }

\newif\if@simboli
\newif\if@riferimenti

\newwrite\file@simboli
\def\simboli{
    \immediate\write16{ !!! Genera il file \jobname.SMB }
    \@simbolitrue\immediate\openout\file@simboli=\jobname.smb}

\newwrite\file@ausiliario
\def\riferimentifuturi{
    \immediate\write16{ !!! Genera il file \jobname.AUX }
    \@riferimentitrue\openin1 \jobname.aux
    \ifeof1\relax\else\closein1\relax\input\jobname.aux\fi
    \immediate\openout\file@ausiliario=\jobname.aux}

\newcount\eq@num\global\eq@num=0
\newcount\sect@num\global\sect@num=0

\newif\if@ndoppia
\def\numerazionedoppia{\@ndoppiatrue\gdef\la@sezionecorrente{\the\sect@num}}

\def\se@indefinito#1{\expandafter\ifx\csname#1\endcsname\relax}
\def\spo@glia#1>{} 

\newif\if@primasezione
\@primasezionetrue

\def\s@ection#1\par{\immediate
    \write16{#1}\if@primasezione\global\@primasezionefalse\else\goodbreak
    \vskip\spaziosoprasez\fi\noindent
    {\bf#1}\nobreak\vskip\spaziosottosez\nobreak\noindent}
%

\def\sezpreset#1{\global\sect@num=#1
    \immediate\write16{ !!! sez-preset = #1 }   }

\def\spaziosoprasez{50pt plus 60pt}
\def\spaziosottosez{15pt}

\def\sref#1{\se@indefinito{@s@#1}\immediate\write16{ ??? \string\sref{#1}
    non definita !!!}
    \expandafter\xdef\csname @s@#1\endcsname{??}\fi\csname @s@#1\endcsname}

\def\adv#1{\global\advance\sect@num by #1}

\def\autosez#1#2#3\par{
    \global\advance\sect@num by 1\if@ndoppia\global\eq@num=0\fi
    \xdef\la@sezionecorrente{\the\sect@num}
    \def\usa@getta{1}\se@indefinito{@s@#1}\def\usa@getta{2}\fi
    \expandafter\ifx\csname @s@#1\endcsname\la@sezionecorrente\def
    \usa@getta{2}\fi
    \ifodd\usa@getta\immediate\write16
      { ??? possibili riferimenti errati a \string\sref{#1} !!!}\fi
    \expandafter\xdef\csname @s@#1\endcsname{\la@sezionecorrente}
    \immediate\write16{\la@sezionecorrente. #2}
    \if@simboli
      \immediate\write\file@simboli{ }\immediate\write\file@simboli{ }
      \immediate\write\file@simboli{  Sezione
                                  \la@sezionecorrente :   sref.   #1}
      \immediate\write\file@simboli{ } \fi
    \if@riferimenti
      \immediate\write\file@ausiliario{\string\expandafter\string\edef
      \string\csname\b@lank @s@#1\string\endcsname{\la@sezionecorrente}}\fi
    \goodbreak\vskip 48pt plus 60pt
\centerline{\lltitle #2}                     
\centerline{\lltitle #3}                     
\par\nobreak\vskip 15pt \nobreak\noindent}

\def\semiautosez#1#2\par{
    \gdef\la@sezionecorrente{#1}\if@ndoppia\global\eq@num=0\fi
    \if@simboli
      \immediate\write\file@simboli{ }\immediate\write\file@simboli{ }
      \immediate\write\file@simboli{  Sezione ** : sref.
          \expandafter\spo@glia\meaning\la@sezionecorrente}
      \immediate\write\file@simboli{ }\fi
\noindent\lltitle \s@ection#2 \par}


\def\eqpreset#1{\global\eq@num=#1
     \immediate\write16{ !!! eq-preset = #1 }     }

\def\eqref#1{\se@indefinito{@eq@#1}
    \immediate\write16{ ??? \string\eqref{#1} non definita !!!}
    \expandafter\xdef\csname @eq@#1\endcsname{??}
    \fi\csname @eq@#1\endcsname}

\def\eqlabel#1{\global\advance\eq@num by 1
    \if@ndoppia\xdef\il@numero{\la@sezionecorrente.\the\eq@num}
       \else\xdef\il@numero{\the\eq@num}\fi
    \def\usa@getta{1}\se@indefinito{@eq@#1}\def\usa@getta{2}\fi
    \expandafter\ifx\csname @eq@#1\endcsname\il@numero\def\usa@getta{2}\fi
    \ifodd\usa@getta\immediate\write16
       { ??? possibili riferimenti errati a \string\eqref{#1} !!!}\fi
    \expandafter\xdef\csname @eq@#1\endcsname{\il@numero}
    \if@ndoppia
       \def\usa@getta{\expandafter\spo@glia\meaning
       \la@sezionecorrente.\the\eq@num}
       \else\def\usa@getta{\the\eq@num}\fi
    \if@simboli
       \immediate\write\file@simboli{  Equazione
            \usa@getta :  eqref.   #1}\fi
    \if@riferimenti
       \immediate\write\file@ausiliario{\string\expandafter\string\edef
       \string\csname\b@lank @eq@#1\string\endcsname{\usa@getta}}\fi}

\def\autoeqno#1{\eqlabel{#1}\eqno(\csname @eq@#1\endcsname)}
\def\autoleqno#1{\eqlabel{#1}\leqno(\csname @eq@#1\endcsname)}
\def\eqrefp#1{(\eqref{#1})}


\def\eq{\autoeqno}
\def\req{\eqrefp}
\def\chap{\autosez}        
\def\nochap{\semiautosez}  



\newcount\cit@num\global\cit@num=0

\newwrite\file@bibliografia
\newif\if@bibliografia
\@bibliografiafalse

\def\lp@cite{[}
\def\rp@cite{]}
\def\trap@cite#1{\lp@cite #1\rp@cite}
\def\lp@bibl{[}
\def\rp@bibl{]}
\def\trap@bibl#1{\lp@bibl #1\rp@bibl}

\def\refe@renza#1{\if@bibliografia\immediate        
    \write\file@bibliografia{
    \string\item{\trap@bibl{\cref{#1}}}\string
    \bibl@ref{#1}\string\bibl@skip}\fi}

\def\ref@ridefinita#1{\if@bibliografia\immediate\write\file@bibliografia{
    \string\item{?? \trap@bibl{\cref{#1}}} ??? tentativo di ridefinire la
      citazione #1 !!! \string\bibl@skip}\fi}

\def\bibl@ref#1{\se@indefinito{@ref@#1}\immediate
    \write16{ ??? biblitem #1 indefinito !!!}\expandafter\xdef
    \csname @ref@#1\endcsname{ ??}\fi\csname @ref@#1\endcsname}

\def\c@label#1{\global\advance\cit@num by 1\xdef            
   \la@citazione{\the\cit@num}\expandafter
   \xdef\csname @c@#1\endcsname{\la@citazione}}

\def\bibl@skip{\vskip +4truept}


\def\stileincite#1#2{\global\def\lp@cite{#1}\global   
    \def\rp@cite{#2}}                                 
\def\stileinbibl#1#2{\global\def\lp@bibl{#1}\global   
    \def\rp@bibl{#2}}                                 

\def\citpreset#1{\global\cit@num=#1
    \immediate\write16{ !!! cit-preset = #1 }    }

\def\autobibliografia{\global\@bibliografiatrue\immediate
    \write16{ !!! Genera il file \jobname.BIB}\immediate
    \openout\file@bibliografia=\jobname.bib}

\def\cref#1{\se@indefinito                  
   {@c@#1}\c@label{#1}\refe@renza{#1}\fi\csname @c@#1\endcsname}

\def\cite#1{\trap@cite{\cref{#1}}}                  
\def\ccite#1#2{\trap@cite{\cref{#1},\cref{#2}}}     
\def\ncite#1#2{\trap@cite{\cref{#1}--\cref{#2}}}    
\def\upcite#1{$^{\,\trap@cite{\cref{#1}}}$}               
\def\upccite#1#2{$^{\,\trap@cite{\cref{#1},\cref{#2}}}$}  
\def\upncite#1#2{$^{\,\trap@cite{\cref{#1}-\cref{#2}}}$}  
\def\Rcite#1{Ref. \cref{#1}}
\def\Rccite#1#2{Refs. \cref{#1},\cref{#2} }
\def\clabel#1{\se@indefinito{@c@#1}\c@label           
    {#1}\refe@renza{#1}\else\c@label{#1}\ref@ridefinita{#1}\fi}
\def\cclabel#1#2{\clabel{#1}\clabel{#2}}                     
\def\ccclabel#1#2#3{\clabel{#1}\clabel{#2}\clabel{#3}}       

\def\biblskip#1{\def\bibl@skip{\vskip #1}}           

\def\insertbibliografia{\if@bibliografia             
    \immediate\write\file@bibliografia{ }
    \immediate\closeout\file@bibliografia
    \catcode`@=11\input\jobname.bib\catcode`@=12\fi}


\def\commento#1{\relax}
\def\biblitem#1#2\par{\expandafter\xdef\csname @ref@#1\endcsname{#2}}


\catcode`@=12


\magnification=1200
\topskip 20pt
\def\interlinea{\baselineskip=16pt}
\def\standardpage{\vsize=21.0truecm\voffset=+0.8truecm
                  \hsize=15.truecm\hoffset=+10truemm
                  \parindent=1.2truecm}

\tolerance 100000
\biblskip{+8truept}                        
\def\hbup{\hfill\break\baselineskip 16pt}  


\global\newcount\notenumber \global\notenumber=0
\def\note #1 {\global\advance\notenumber by1 \baselineskip 10pt
              \footnote{$^{\the\notenumber}$}{\nine #1} \interlinea}
\def\clearnotenumber{\notenumber=0}



\font\text=cmr10
\font\scal=cmsy5
\font\bcal=cmsy10 scaled \magstep3   
\font\it=cmti10
\font\name=cmssi10 scaled \magstep1
\font\lname=cmssi9
\font\btitle=cmbx10 scaled \magstep4     
\font\title=cmbx10 scaled \magstep3      
\font\ltitle=cmbx12 scaled \magstep1
\font\lltitle=cmbx12
\font\bmtitle=cmmi12                      
\font\mtitle=cmmi12 
\font\lmtitle=cmmi12                     
\font\itltitle=cmti10 scaled \magstep2   
\font\ittitle=cmti10 scaled \magstep3    
\font\subtitle=cmbx10 scaled \magstep1
\font\abs=cmti10 scaled \magstep1        

\font\mb=cmmib10
\font\seven=cmr7                         
\font\eight=cmr8                         
\font\nine=cmr9                         
\font\bcp=cmbx7
\font\icp=cmmi7
\font\bb=cmbx12 scaled \magstep1
\font\bl=cmbx12








\def\gtrsim{\ \rlap{\raise 2pt \hbox{$>$}}{\lower 2pt \hbox{$\sim$}}\ }
\def\lesssim{\ \rlap{\raise 2pt \hbox{$<$}}{\lower 2pt \hbox{$\sim$}}\ }


\def\salta{\vskip 2pt}
\def\mn{\medskip\noindent}
\def\bs{\bigskip}
\def\hb{\hfil\break}
\def\no{\noindent}

\def\scs{\scriptstyle}
\def\scss{\scriptscriptstyle}
\def\ph#1{\phantom#1}
\def\o{\over}



\def\ea{{\it et.al.}}
\def\ib{{\it ibid.\ }}

\def\npb#1{Nucl. Phys. {\bf B#1},}
\def\plb#1{Phys. Lett. {\bf B#1},}
\def\prd#1{Phys. Rev. {\bf D#1},}
\def\prl#1{Phys. Rev. Lett. {\bf #1},}
\def\ncim#1{Nuo. Cim. {\bf #1},}
\def\zpc#1{Z. Phys. {\bf C#1},}
\def\prep#1{Phys. Rep. {\bf #1},}
\def\rmp{Rev. Mod. Phys. }
\def\ijmpa#1{Int. J. Mod. Phys. {\bf A #1},}


\stileincite{}{}     
\numerazionedoppia   

\interlinea
\standardpage
\text                


\def\scs#1{{\scriptstyle #1}}
\def\scss#1{{\scriptscriptstyle #1}}
\def\ph#1{\phantom{#1}}
\def\o{\over}
\def\half{{1\over 2}}


\def\clel{c_L^{e}}
\def\slel{s_L^{e}}
\def\crel{c_R^{e}}
\def\srel{s_R^{e}}

\def\clmu{c_L^{\mu}}
\def\slmu{s_L^{\mu}}
\def\crmu{c_R^{\mu}}
\def\srmu{s_R^{\mu}}

\def\cltau{c_L^{\tau}}
\def\sltau{s_L^{\tau}}
\def\crtau{c_R^{\tau}}
\def\srtau{s_R^{\tau}}

\def\clu{c_L^{u}}
\def\slu{s_L^{u}}
\def\cru{c_R^{u}}
\def\sru{s_R^{u}}

\def\cld{c_L^{d}}
\def\sld{s_L^{d}}
\def\crd{c_R^{d}}
\def\srd{s_R^{d}}

\def\cls{c_L^{s}}
\def\sls{s_L^{s}}
\def\crs{c_R^{s}}
\def\srs{s_R^{s}}

\def\clc{c_L^{c}}
\def\slc{s_L^{c}}
\def\crc{c_R^{c}}
\def\src{s_R^{c}}

\def\clb{c_L^{b}}
\def\slb{s_L^{b}}
\def\crb{c_R^{b}}
\def\srb{s_R^{b}}

\def\clnue{c_L^{\nu_e}}
\def\slnue{s_L^{\nu_e}}
\def\crnue{c_R^{\nu_e}}
\def\srnue{s_R^{\nu_e}}

\def\clnumu{c_L^{\nu_\mu}}
\def\slnumu{s_L^{\nu_\mu}}
\def\crnumu{c_R^{\nu_\mu}}
\def\srnumu{s_R^{\nu_\mu}}

\def\clnutau{c_L^{\nu_\tau}}
\def\slnutau{s_L^{\nu_\tau}}
\def\crnutau{c_R^{\nu_\tau}}
\def\srnutau{s_R^{\nu_\tau}}

\def\kud{\kappa_{ud}}
\def\kus{\kappa_{us}}
\def\kcd{\kappa_{cd}}
\def\kcs{\kappa_{cs}}


\def\A{{\cal A}}
\def\ww{M^2_W}
\def\zz{M^2_Z}
\def\zzo{M^2_{Z_0}}
\def\zzp{M^2_{Z^\prime}}
\def\ss{s^2_w}
\def\cc{c^2_w}
\def\G{{\cal G_{\rm SM}}}
\def\E{{\rm E}_6}
\def\sb{s_{\beta}}
\def\cb{c_{\beta}}

\def\N{{\hbox{\scal N}}}
\def\K{{\hbox{\scal K}}}


\def\pr{\prime}
\def\nubar{{\buildrel (-) \over \nu}}
\def\numubar{{\buildrel (-) \over {\nu_\mu}}}


\font\mbf=cmmib10  scaled \magstep 2      
\def\bfchi{{\hbox{\mbf\char'037}}}
\def\bftau{{\hbox{\mbf\char'034}}}
\def\bflambda{{\hbox{\mbf\char'025}}}
\def\bfmu{{\hbox{\mbf\char'026}}}
\def\bfepsilon{{\hbox{\mbf\char'017}}}
\def\bfe{{\hbox{\mbf\char'145}}}
\def\bfu{{\hbox{\mbf\char'165}}}
\def\bfF{{\hbox{\mbf\char'106}}}
\def\bfC{{\hbox{\mbf\char'103}}}
\def\bfS{{\hbox{\mbf\char'123}}}
\def\bfa{{\hbox{\mbf\char'141}}}

\font\bigm=cmmi12            
\def\bchi{{\hbox{\bigm\char'037}}}
\def\btau{{\hbox{\bigm\char'034}}}
\def\blambda{{\hbox{\bigm\char'025}}}
\def\bmu{{\hbox{\bigm\char'026}}}
\def\bepsilon{{\hbox{\bigm\char'017}}}
\def\be{{\hbox{\bigm\char'145}}}
\def\bu{{\hbox{\bigm\char'165}}}
\def\bF{{\hbox{\bigm\char'106}}}
\def\bC{{\hbox{\bigm\char'103}}}
\def\bS{{\hbox{\bigm\char'123}}}
\def\ba{{\hbox{\bigm\char'141}}}

\def\ie{{\it i.e.\ }}
\def\eg{{\it e.g.\ }}

\def\ra{\rangle}
\def\la{\langle}

\def\scs{\scriptstyle }
\def\scss{\scriptscriptstyle }
\def\ph#1{\phantom{(#1)}}
\def\o{\over}
\def\half{{1\over 2}}


\def\lu{{H^cQu^c}}
\def\ld{{HQd^c}}
\def\lt{{HLe^c}}
\def\lq{{S^chh^c}}
\def\lc{{hu^ce^c}}
\def\ls{{LQh^c}}
\def\lst{{\nu^chd^c}}
\def\lo{{hQQ}}
\def\ln{{h^cu^cd^c}}
\def\ldi{{H^cL\nu^c}}
\def\luu{{H^cHS^c}}


\def\l#1#2#3#4{\lambda^{^{\scriptstyle (#1)}}_{\scriptscriptstyle #2#3#4}\>}
\def\lw#1{\lambda^{^{\scriptstyle (#1)}}\>}

\def\ww{M^2_W}
\def\zz{M^2_Z}
\def\zzo{M^2_{Z_0}}
\def\zzu{M^2_{Z_1}}
\def\zzp{M^2_{Z^\prime}}
\def\G{{\cal G_{\rm SM}}}
\def\E{{\rm E}_6}
\def\sb{s_{\beta}}
\def\cb{c_{\beta}}

\def\pr{\prime}
\def\nubar{{\buildrel (-) \over \nu}}
\def\numubar{{\buildrel (-) \over {\nu_\mu}}}

\font\mbf=cmmib10  scaled \magstep1      
\def\bfchi{{\hbox{\mbf\char'037}}}
\def\bftau{{\hbox{\mbf\char'034}\>}}
\def\bflambda{{\hbox{\mbf\char'025}}}
\def\bfmu{{\hbox{\mbf\char'026}}}
\def\bfepsilon{{\hbox{\mbf\char'017}}}
\def\bfe{{\hbox{\mbf\char'145}}}
\def\bfu{{\hbox{\mbf\char'165}}}
\def\bfF{{\hbox{\mbf\char'106}}}
\def\bfC{{\hbox{\mbf\char'103}}}
\def\bfS{{\hbox{\mbf\char'123}}}
\def\bfa{{\hbox{\mbf\char'141}}}

\font\bigm=cmmi12            
\def\bchi{{\hbox{\bigm\char'037}}}
\def\btau{{\hbox{\bigm\char'034}\>}}
\def\blambda{{\hbox{\bigm\char'025}}}
\def\bmu{{\hbox{\bigm\char'026}}}
\def\bepsilon{{\hbox{\bigm\char'017}}}
\def\be{{\hbox{\bigm\char'145}}}
\def\bu{{\hbox{\bigm\char'165}}}
\def\bF{{\hbox{\bigm\char'106}}}
\def\bC{{\hbox{\bigm\char'103}}}
\def\bS{{\hbox{\bigm\char'123}}}
\def\ba{{\hbox{\bigm\char'141}}}

\autobibliografia
\pageno=0
\vsize=23.8truecm
\hsize=15.7truecm
\voffset=-1.truecm
\hoffset=+6truemm
\baselineskip 12pt
\rightline{UM-TH 93--19}\par\noindent
\rightline{hep-ph/9309239}\par\noindent

\bs\bs\bs
\centerline{\ltitle
Signals of Unconventional E$_{\bf 6}$ Models}
\bs
\centerline{\ltitle at \bfe$^{\displaystyle\bf +}$
\bfe$^{\displaystyle\bf -}$ Colliders.}
\medskip
\medskip
\bs\bs                                     
                 \centerline{Enrico Nardi}
\bs
\centerline{\it Randall Laboratory of Physics, University of Michigan,
           Ann Arbor, MI 48109--1120}
\vskip 1truecm                             

\medskip
\centerline{\bf {\abs Abstract}}           
\bs

\noindent
Generation dependent discrete symmetries
often appear in models derived from superstring theories.
In particular, in the framework of $\E$ models the presence of such
symmetries is required in order to allow for the radiative generation
of naturally small neutrino masses.
Recently it was shown that
by imposing suitable generation dependent
discrete symmetries, a class of models  can be consistently
constructed in which  the three sets of known fermions
in each generation do not have the same assignments with
respect to the {\bf 27} representation of E$_6$.
In this scenario, the different embedding in the gauge group of the
three generations implies in particular
that the known charged leptons couple in a
non--universal way to the new neutral gauge bosons $(Z_\beta)$
present in these models.
We exploit this fact to study the signature of this class of models
at present and future $e^+e^-$ colliders. We show that
some signals of deviation from lepton universality
as well as  some other discrepancies with the standard model predictions
which have been observed at the
TRISTAN collider in the production rate of $\mu$ and $\tau$,
can be accounted for if the $Z_\beta$ mass is not
much heavier than 300 GeV. We also study the discovery limits for
lepton universality violation of this type at LEP-2 and at the 500
GeV $e^+e^-$ Next Linear Collider (NLC). We show that models
predicting unconventional assignments for the leptons will give an
unmistakable signature, when the $Z_\beta$ mass is as heavy as $\sim 800$
GeV (LEP-2) and $\sim 2$ TeV (NLC).
\bs
\noindent
PACS number(s): 12.10.Dm,12.15.Ff,13.15.Jr,14.60.Gh
\vfill
\noindent
--------------------------------------------\phantom{-} \hb
\leftline{Electronic address: nardi@umiphys.bitnet}
\bigskip
\leftline{UM-TH 93--19}
                   \medskip
\centerline{August 1993}

\eject

\standardpage                            
\interlinea                              

\null
\chap{Int} {I. INTRODUCTION}{}

It was recently shown [\cite{f}] that in the framework of
superstring derived E$_{\bf 6}$ models,
it is possible to implement an
unconventional scenario in which
some of the known fermions of the three families are embedded in the
fundamental {\bf 27} representation of the group in a generation
dependent way, meaning that their gauge quantum numbers do not
replicate throughout the three generations.
It was also argued [\cite{f}] that if
$\E$ models are required to allow for small
neutrino  masses (as is needed for any
particle physics solution of the solar neutrino problem,
and/or for explaining the  atmospheric $\nu_\mu$ deficit via
$\nu_\mu$ oscillations)
the Unconventional Assignments (UA) scenario should
be considered as a natural alternative to the standard schemes.
In fact, in the framework of superstring inspired $\E$ theories
the most  natural way for generating small neutrino masses
is through radiative corrections [\cite{ellis-e6},\cite{MNS}]
since in these models the Higgs representation necessary to implement
a see-saw mechanism [\cite{see-saw}]
is absent.\note{\noindent See however Ref. [\cite{nandi-sarkar}]
for a discussion of a see-saw mechanism induced by gravitational effects.}
In order to implement the generation of $\nu$ masses at the loop
level, a  suitable discrete symmetry must be imposed on
the superpotential to insure that at the lowest order $m_\nu=0$.
Branco and Geng [\cite{branco}] have shown that no
generation-blind symmetry exists that forbids non--vanishing neutrino
masses at the tree level and at the same time allows for their
radiative generation. As a result, in order to implement in a consistent way
the generation of $\nu$ masses via loop diagrams, a symmetry that
does not act in the same way on the three generations is needed.

The main motivation for investigating the UA schemes stems from the
observation that once we chose to build a model based on a symmetry that
distinguishes among the different generations, there is no reason in
principle to expect that this symmetry will result in a set of {\it
light} fermions  (\ie the known states) that will exactly replicate
throughout the three generations [\cite{f}]. This implies the possibility
that the known states belonging to different generations could
have different $\E$ gauge interactions.
Of course, experimentally  we know that the $SU(2)\times U(1)$ gauge
interactions of the known fermions do respect universality with a high
degree of precision. However, since the Standard Model (SM)
gauge group is rank 4, while  $\E$ is rank  6, as many as two
additional massive neutral gauge bosons $(Z_\beta)$ can be present,
possibly with $M_\beta \sim 1\>$TeV or less, and the possibility that
the additional $U_\beta(1)$ interactions could violate universality is
still phenomenologically viable.

Since the fundamental representation of $\E$ is 27 dimensional,
the fermion content of models based on this group
is enlarged with respect to the SM.
In fact, in addition to the standard fermions,
two additional leptonic $SU(2)$-doublets, two $SU(2)$-singlet
neutral states and two color-triplet $SU(2)$-singlet $d$-type quarks
are present. UA models are realised by identifying
in a generation dependent way some of the
known doublets of lepton and/or $d$-type singlet quarks
with the additional fermion multiplets [\cite{f}].
Models based on the UA scenario can be implemented without conflicting
with phenomenological or theoretical constraints. For example the
model described in Ref. [\cite{f}] was shown to be consistent with a
large number of experimental constraints, ranging from the direct and
cosmological limits on the neutrino masses, to the stringent limits on
flavor changing neutral currents (FCNCs). In this model the left (L)
handed lepton doublet ``${\nu_\tau \choose \tau}_L$'' and the right
(R) handed quark singlet ``$b_R$'' of the third
generation are assigned to $SU(2)$ multiplets having a different
embedding in $\E$ with respect to the
corresponding states of the first two generations. The non standard
phenomenology resulting from this model implies in particular that the
``$\tau$" neutrinos have different neutral current
(NC) interactions than to ``$\nu_e$'' and ``$\nu_\mu$''.

In Sec. 2 we will briefly outline the main features of the $\E$ models
based on the UA scenario, and establish our conventions
and notations. A more complete discussion of the theoretical framework
can be found in
Ref. [\cite{f}].  A very
clean signature for the UA models would be the detection of deviations
from universality in neutral current (NC) processes.
Due to the clean experimental environment of $e^+e^-$ annihilation,
such a signature could be more easily detected at $e^+e^-$ colliders
rather than at hadron colliders.
In Sec. 3 we will investigate the phenomenology of UA models
at the present and future $e^+e^-$ machines.
Since
the UA for the known fermions would result in a violation
of universality only in the fermion couplings to the new $Z_\beta$
bosons without affecting the couplings to the standard $Z_0$,
the large amount of data collected
at the $Z_0$ resonance by the Large Electron Positron (LEP) collaborations
are not effective for the search of these kind of effects.
In fact, the contribution of $Z_\beta$--$Z_0$ interference
to the various cross sections and asymmetries measured at LEP-1
vanishes at the peak, and the contribution of pure $Z_\beta$ exchange
is also vanishingly small.
Some effects could still be detected
if the $Z_0$ had a sizeable mixing with $Z_\beta$,
however the existing bounds  on the  $Z_0$--$Z_\beta$
mixing angle are extremely tight [\cite{zp-new}], so that
we will disregard this possibility throughout this paper.

Among presently operating colliders, the one best suited
to reveal the kind of effects we are looking for is
the TRISTAN collider at the KEK laboratories, which is collecting data
at about 60 GeV c.m. energy.
It is intriguing, though certainly not compelling,  that
a few discrepancies between the
TRISTAN data on the total hadronic and leptonic cross sections
and the SM predictions do exist [\cite{kek}], and they
point towards the existence of a $Z_\beta$ at rather low energies
$M_\beta\lesssim 300\,$ GeV [\cite{zp-kek}].
At the same time the data on the rate of production of
$\tau$ and $\mu$ leptons do show a signal of violation of universality
(at the level of 1.6 standard deviations) which could not be explained
by conventional extended gauge models. However, as we will show,
these data can be well accounted for in the framework of models with UA.
We will complete our discussion by analyzing
the discovery limits for UA models at
LEP II and at the 500 GeV Next Linear $e^+e^-$ Collider (NLC).
Finally, in Sec. 4 we will summarize our results and draw the
conclusions.

\chap{flip} {II. UNCONVENTIONAL ASSIGNMENTS}{IN E$_{\bf 6}$ MODELS}

In $\E$ grand unified theories,
matter fields belong to the fundamental {\bf 27}
representation of the group. $\E$ contains $SO(10)\times U_\psi(1)$ as
a maximal subalgebra, and the {\bf 27} branches to
the ${\bf 1} + {\bf 10}
+ {\bf 16}$ representation of $SO(10)$.
In turn $SU(10)$ contains $SU(5)\times U_\chi(1)$.
The $SO(10)$,  $SU(5)$, $U_\psi(1)$ and $U_\chi(1)$ assignments for
the fermions in the {\bf 27} representation
are listed in Table I. Usually the known
particles of the three generations are assigned to the {\bf 16} of
$SO(10)$ that also contains a $SU(2)$ singlet neutrino ``$\nu^c$"
$$
[{\bf 16}]_i = \big[
Q\equiv { 
{u\choose d}}, \,
u^c, \,
e^c, \,
d^c, \,
L\equiv { 
{\nu \choose e}}, \,
\nu^c\big]_i \qquad \qquad i=1,2,3.
\eq{2.1}
$$
\mn
The {\bf 10} and the {\bf 1} of $SO(10)$
contain the new fields
$$
\eqalign{
[{\bf 10}]_i &= \big[
H^c\equiv {  
{E^c\choose N^c}}, \,
h, \,
H\equiv {  
{N\choose E}}, \,
h^c\big]_i                                          \cr
[\>{\bf 1}\>]_i&= [S^c]_i   \qquad \qquad \qquad \qquad
\hskip 2.8truecm i=1,2,3. \cr}
\eq{2.2}
$$


\def \tabrule     {\noalign{\vskip 5truept \hrule\vskip 5truept} }
\def \tabrul2     {\noalign{\vskip 5truept \hrule \vskip 2truept \hrule
                   \vskip 5truept} }
\def\om{\omit}

\baselineskip=14pt
\midinsert
{
\bs
$$
\vbox{\hsize= 13.3truecm
\noindent TABLE I.
{\nine
$SO(10)$, $U_\psi(1)$, $SU(5)$ and $U_\chi(1)$ assignments for the
left-handed fermions of the {\bf 27} fundamental representation of
${\rm E}_6$. The $SU(2)$ doublets $H^c$, $H$, $L$ and $Q$ are
explicitly written in components. The Abelian charges $Q_\psi$ and
$Q_\chi$ can be obtained from the quantum numbers in the
brackets by dividing
by $c_\psi = 6\sqrt{2/5}$ and $c_\chi =
6\sqrt{2/3}$, respectively.
The charges are normalized to the
hypercharge according to: $\sum_{f=1}^{27}(Q_{\psi,\chi}^f)^2 =
\sum_{f=1}^{27}({1 \over 2}Y^f)^2=5$.
}
\vskip -.8truecm
}
$$

$$
\vbox{
\hsize= 13.35truecm
\offinterlineskip
\halign {
\vrule#&\strut#&\vrule#&\strut#&\vrule#&\strut\quad#&\vrule#&\strut#&
\vrule#&\strut#&\vrule#&\strut#&\vrule#&\strut\quad#&\vrule#&\strut#&
\vrule#&\strut#&\vrule#&\strut#&\vrule#&\strut#&\vrule#&\strut#&\vrule#\cr
\noalign{\hrule}
height6pt & \om && \om && \om&\om&\om && \om&\om&\om && \om &&
\om&\om&\om && \om&\om&\om&\om&\om &\cr
&\om && $\ S^c\> $ && ${{E^c}\choose{N^c}}\ $&\om&$\ h\ $ && $\
{N\choose{E}}\ $&\om&$\ h^c\ \ $ && $\nu^c\ $ && $\ \>
{\nu\choose{e}}$&\om&$\ d^c\ \>$ && $\ \> e^c\ \>$&\om&$\ u^c\
\>$&\om&$\ {u\choose d}\ $ &\cr
height6pt&\om&&\om&&\om&\om&\om&&\om&\om&\om&&\om&&
\om&\om&\om&&\om&\om&\om&\om&\om&\cr
\noalign{\hrule}
height6pt&\om&&\om&&\om&\om&\om&\om&\om&\om&\om&&
\om&\om&\om&\om&\om&\om&\om&\om&\om&\om &\om&\cr
& $\ SO(10)\> (c_\psi Q_\psi)\ $ && $\ {\bf 1}\>(4)\ $ &&   
\multispan7 \hfil ${\bf 10}\> (-2)$\hfil          
&&                                                
\multispan{11} \hfil ${\bf 16}\> (1)$ \hfil       
&\cr
height6pt&\om&&\om&&
\om&\om&\om&\om&\om&\om&\om&&
\om&\om&\om&\om&\om&\om&\om&\om&\om&\om&\om&\cr
\noalign{\hrule}
height6pt&\om&&\om&&\om&\om&\om&&\om&\om&\om&&
\om&&\om&\om&\om&&\om&\om&\om&\om&\om&\cr
& $\ SU(5)\ (c_\chi Q_\chi)\  $ && $\ {\bf 1}\>(0)\ $ &
&\multispan3\hfil ${\bf 5}\>(2)$ \hfil           
&& \multispan3 \hfil ${\bf
\bar 5}\>(-2)$\hfil                  
&&\hfil ${\bf 1}\>(-5)\ $ \hfil
&&                                   
\multispan3 \hfil ${\bf \bar 5}\>(3)$\hfil        
&&                                                
\multispan5
\hfil ${\bf 10}\> (-1)$\hfil                      
&\cr
height6pt & \om && \om && \om&\om&\om && \om&\om&\om && \om &&
\om&\om&\om && \om&\om&\om&\om&\om &\cr
\noalign{\hrule}}}
$$
\bs
}
\vskip -.6truecm
\endinsert
\interlinea

As it is clear from Table I,
there is an ambiguity in assigning the known states to the {\bf 27}
representation, since under the SM gauge group
$$
\G\equiv SU(3)_c\times SU(2)_L\times U(1)_Y
$$
the ${\bf \bar 5}_{({\bf 10})}$ in
the {\bf 10} of $SO(10)$ has the same field content as the ${\bf
\bar 5}_{({\bf 16})}$ in the {\bf 16}. The same ambiguity is
also present for the two $\G$ singlets, namely
 ${\bf 1}_{({\bf 1})}$ and ${\bf 1}_{({\bf 16})}$.
In the present paper we will concentrate on the consequences of
having different assignments for the known L-handed leptons.
Since these leptons might not correspond to the entries as listed in Table I,
we use quotation marks to denote the known states with
their  conventional labels, while labels not enclosed within quotation
marks will always refer to the fields listed in the table.

In the models under investigation
as many as two additional neutral gauge bosons
can be present, corresponding for example, to some linear combinations
of the $U_\chi(1)$ and $U_\psi(1)$ generators.
The interaction of the fermions in the ${\bf \bar 5}$ of $SU(5)$
with these gauge bosons will depend on the specific assignments
to the {\bf 16} or to the {\bf 10} of $SO(10)$.
The
two additional neutral gauge bosons are usually parametrized as
$$
\eqalign{
&Z_\beta^\pr= Z_\psi \sin\beta + Z_\chi \cos\beta  \cr
&Z_\beta^{\pr\pr}= Z_\psi \cos\beta - Z_\chi \sin\beta, \cr}
\eq{2.3}
$$
\mn
where $\beta$ is a model dependent parameter.
In the following we will denote the
lightest of the two new gauge bosons as $Z_\beta$.
In the presence of a `light' $Z_\beta$
different assignments will lead to a different phenomenology.
In contrast, in the limit $M_{\beta}\rightarrow \infty$
the choice of the assignment is irrelevant as long as
we are only concerned with the gauge interactions.  However,
even in
this limit the requirement of $U_{\psi,\chi}(1)$ gauge invariance for the
superpotential, together with the phenomenological constraints on the
absence of FCNCs in the Higgs sector, strongly
constrain the structure of the viable models [\cite{f}].

A model realizing an UA scenario,
in which what we call ``$\tau_L$" corresponds
to the charged component of the $H_3$ weak doublet belonging to
${\bf \bar 5}_{{\bf 10}}$, while the
``$e_L$'' and the ``$\mu_L$'' leptons are  as usual assigned to the
${\bf \bar 5}_{{\bf 16}}$, was recently proposed in Ref. [\cite{f}].
This model is realized by imposing on the superpotential
a particular family-non-blind
$Z_2\times Z_3$ discrete symmetry.
As a result  of such a symmetry,
the masses of the known (light) chiral leptons
are generated by vacuum expectation values (VEVs) of Higgs doublets,
through the terms
$m_\tau E_{3L}e_{3R}$
(with $m_\tau\sim \la\tilde L_3\ra_0$) and
$m_{\alpha\beta}e_{\alpha L}e_{\beta R}$
(with $m_{\alpha\beta} \sim \la\tilde H_2\ra_0$ and
$\alpha,\beta=1,2$).
The remaining charged leptons
$e_{3L}$, $E_{3R}$,
$E_{\alpha L}$, $E_{\alpha R}$ are vectorlike, and acquire large
masses from VEVs of Higgs singlets.

As it was argued in Ref. [\cite{f}], several interesting features
of this model are peculiar to the UA schemes in general,
independently of this particular realization.
For example, in contrast to the conventional $\E$ models  [\cite{ellis-e6}],
in the UA schemes
rank 6 models are not disfavoured with respect to rank 5,
so that the general parametrization of the two additional gauge bosons
given in \req{2.3} is well motivated.
We stress that other assignments, leading to models with a structure
similar to the model proposed in Ref. [\cite{f}], but implying
a different phenomenology, can be easily obtained
by means of some different discrete symmetries.

If some of the
$\nu^c$ and/or $S^c$ $\>SU(2)$ neutral
singlets
are massless or are very light ($m_s\lesssim 1\,$ MeV),
cosmological arguments suggest that the $Z_\beta$
bosons should be heavier than about $\sim 1$-2 TeV
[\cite{nucleo-e6}], thus excluding the possibility
of detecting any signal at TRISTAN and LEP-2.
In fact, though singlet under $\G$, these states do
have $U(1)_\beta$ interactions.
Then, not to conflict with the limit of 3.6 relativistic neutrinos in
thermal equilibrium at the time of nucleosynthesis [\cite{nucleo-nu}]
(which can be derived from the data on the light element abundances) we
have to require this interaction to be weak enough to allow for the
decoupling of the light $\G$-singlets at a sufficiently early time
(for example
before the QCD phase transition) so that their number density can be
safely diluted. Requiring the $U_\beta(1)$
interaction to be ``superweak" results in
the quoted lower bound on the $Z_\beta$ mass [\cite{nucleo-e6}].
We would like to mention that there are two models
corresponding to the particular values of the angle $\beta$
in \req{2.3}
($\tan 2\beta = -{\sqrt{15} \o 7},\> 0$)
in which the
nucleosynthesis constraints on $M_\beta$ can be evaded
even in the presence of light
$SU(2)$ singlets [\cite{f}].
In fact,  for these two values of
$\beta$, respectively
the $\nu^c$ and the $S^c$ degrees of freedom decouple from
the lightest $Z_\beta$, behaving as
`effective singlets' with respect
to all the `light' gauge bosons.
Then they  could play the
role of the helicity partners of the standard
neutrinos, allowing in particular for non zero neutrino Dirac masses,
while at the same time their gauge interactions would not be effective
to keep them in thermal equilibrium in the early Universe.

However, for the sake of generality, in the present analysis we will
assume that all the $\nu^c$ and $S^c\>$ $\G$-singlets
are heavy ($m_s\gg 1\,$ MeV).
In this case,
independently of the value of $\beta$,
the nucleosynthesis constraints on $M_\beta$ are evaded.
Therefore the $Z_\beta$  boson could be as light as
allowed by the present limits from direct searches at colliders
[\cite{zp-direct}] and from the analysis of $Z_\beta$ indirect effects
[\cite{zp-new}], resulting in both cases in $M_\beta \gtrsim
200-300\,$GeV.
As we will see, in the UA schemes
the presence of a $Z_\beta$ with a mass in this range
can give rise to lepton universality violating
effects that could be
detected at the colliders presently in operation.

\chap{phen} {III. SIGNALS OF UNCONVENTIONAL ASSIGNMENTS}
{AT \bfe$^{\displaystyle\bf +}$ \bfe$^{\displaystyle\bf -}$ COLLIDERS}

In the presence of additional neutral gauge bosons, the
lowest order cross section\note{In the numerical computations
we have taken into account the leading one-loop corrections by using an
improved Born approximation [\cite{leprad}].}
for the process $e^+e^-\to l^+l^-$ with $l\neq e$, is
$$
\eqlabel{3.1}
\eqlabel{3.2}
\eqlabel{3.3}
\eqalignno{
\sigma(s)&= {4\o 3}{\pi\alpha^2\o s}\sum_{m,n=0}^N
C(m,n)\chi_m(s)\chi_n^*(s)                      &\req{3.1}  \cr
&\cr
C(m,n)&=
\left[v_m(e)v_n^*(e)+ a_m(e)a_n^*(e) \right]\cdot
\left[v_m(l)v_n^*(l)+ a_m(l)a_n^*(l) \right]    &\req{3.2}   \cr
&\cr
 \chi_m(s)&={g^2_m\o 4\pi\alpha}{s\o s - M^2_m - iM_m\Gamma_m}.
                        &\req{3.3}
 \cr}
$$
\mn
We will henceforth assume
that one of the two new bosons in \req{2.3} is  heavy enough so that
its effects on the low energy physics are negligible. Then
$m,n=0$,$1$,$2$ correspond respectively to the
$\gamma$, $Z_0$ and $Z_\beta$
amplitudes. The couplings in \req{3.2} and \req{3.3} are
$$
\eqlabel{3.4}
\vbox{
\baselineskip=18pt
\settabs \+
$g_1=(\sqrt{2}G_\mu M^2_Z)^{1\o 2}$ \quad
&$v_1(\ell)=T_{3L}^\ell-2Q_{\rm em}^\ell s^2_w$   \quad
&$a_2(\ell)=Q_\beta^\ell+Q_\beta^{\ell^c}$   \quad
&\quad$\ \ell=e,\mu,\tau$
\cr  \+
$g_0=e$
&$v_0(\ell)=Q_{\rm em}^\ell$
&$a_0(\ell)=0$ &
\cr  \+
$g_1=(\sqrt{2}G_\mu M^2_Z)^{1\o 2}$
&$v_1(\ell)=T_{3L}^\ell-2Q_{\rm em}^\ell s^2_w$
&$a_1(\ell)=T_{3L}^\ell$
&  &\phantom{\quad$\ \ell=e,$}\req{3.4}
\cr  \+
$g_2=s_w g_1$
&$v_2(\ell)=Q_\beta^\ell-Q_\beta^{\ell^c}$
&$a_2(\ell)=Q_\beta^\ell+Q_\beta^{\ell^c}$
&\quad$ \ell=e,\mu,\tau$
\cr  }
$$
\interlinea
\mn
where $Q_{\rm em}^\ell=-1$ is the electric charge of the leptons,
$T_{3L}^\ell=-{1\o 2}$ is the third component of the
weak isospin, $s_w\equiv \sin\theta_w$ with $\theta_w$
the weak mixing angle, $Q_\beta^{\ell,\ell^c} =
Q_\psi^{\ell,\ell^c}\sin\beta + Q_\chi^{\ell,\ell^c}\cos\beta\ $
is the lepton coupling to the $Z_\beta$ boson in \req{2.3}.
The new charges
$Q_\psi^{\ell,\ell^c}$, $Q_\chi^{\ell,\ell^c}$ are given in Table I,
and are normalized to the
hypercharge generator $Y/2$. In addition,
in \req{3.4} we have assumed
for the abelian coupling, $g_2$,
a renormalization
group evolution down to the electroweak scale
similar to that of
the hypercharge coupling $g_{\scss Y}\simeq s_w g_1$.

In \req{3.4} the vector and axial-vector couplings,
$v_{0,1}(\ell)$ and $a_{0,1}(\ell)$,
do not  depend on the specific assignments for the leptons, and
are unmodified with respect to the SM.
In contrast, $v_{2}(\ell)$ and $a_{2}(\ell)$
do depend on the particular assignments of the $\ell$ lepton.
With the notations given in \req{2.1} and \req{2.2}, and
referring to the {\bf 16} and to the {\bf 10} representations
of $SO(10)$, the possible assignments for the
L-handed $\ell_1=$``$e_L$'', $\ell_2=$``$\mu_L$'',
$\ell_3=$``$\tau_L$''
charged leptons are
$$
\eqalign{
&``\ell_i" \in L_i \in {\bf 16} \cr
&{\rm or }                       \cr
&``\ell_i" \in H_i \in {\bf 10} \qquad\qquad i=1,2,3.  \cr
}
\eq{3.5}
$$
\mn
With these assignments
three different cross sections for the process
$e^+e^- \to l^+l^-$ ($l\neq e$)
are possible. They are
$\sigma_{16\to 16}$,
$\sigma_{16\to 10} = \sigma_{10\to 16}$ and
$\sigma_{10\to 10}$, where the subscripts refer to
the specific embedding of the
L-handed components of the initial $e^-$ and final
$l^-$ states in the {\bf 16} or in the {\bf 10} of $SO(10)$.
In the following we will give results for the
quantities $R_{16\to 16}$, $R_{16\to 10}$ and $R_{10\to 10}$
corresponding to the different cross sections
normalized to the point-like QED cross section for muon pair
production.

\midinsert
{
\vskip 1truecm
\epsfxsize=14.0truecm 
\epsffile{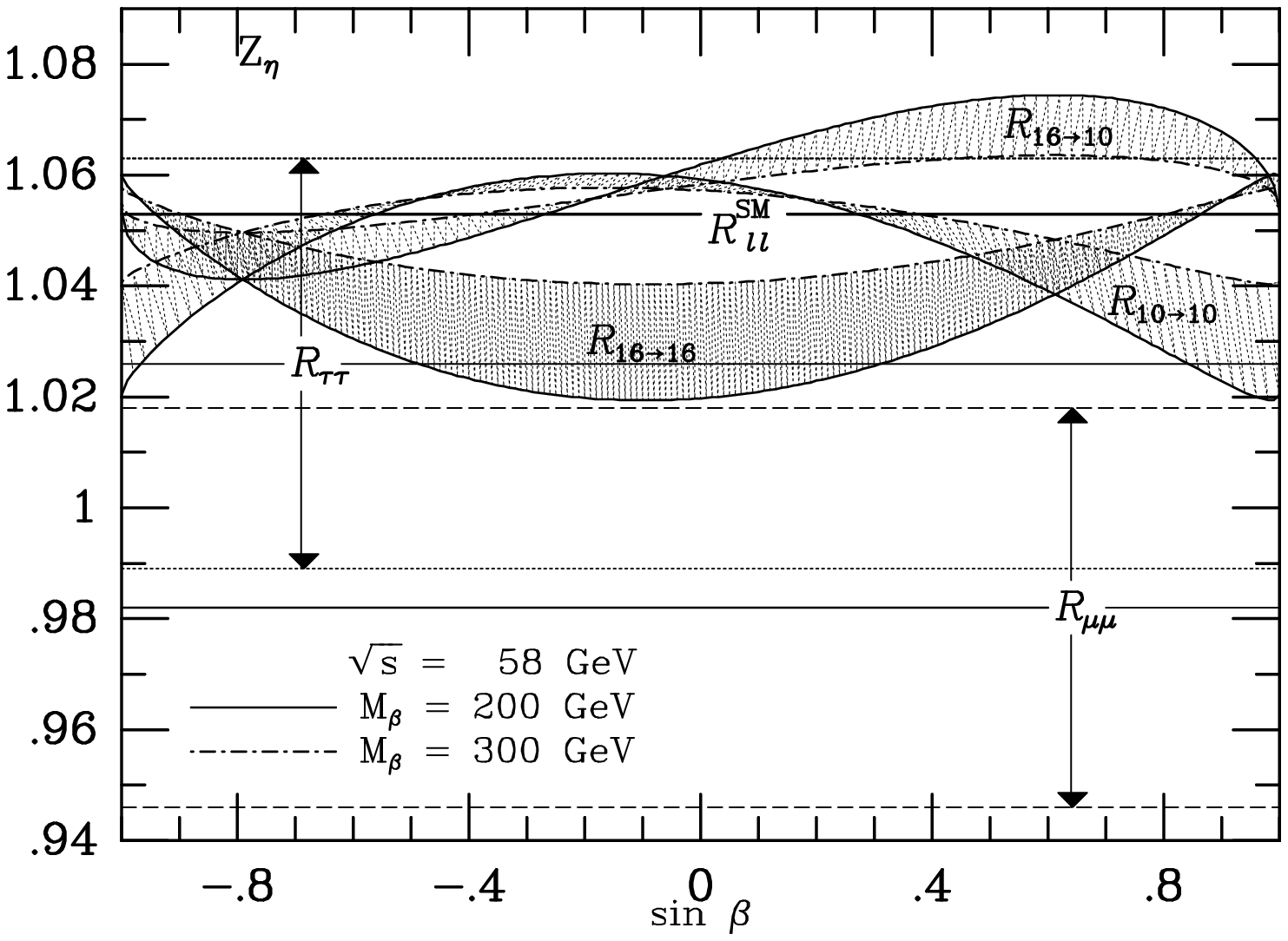}
\vskip -10.5truecm
\baselineskip=12pt
$$
\vbox{
\hsize= 13.3truecm
\noindent
FIG. 1.
{\nine
The TRISTAN total cross sections
 for $\mu$ pair production ($R_{\mu\mu}$) and
$\tau$ pair production ($R_{\tau\tau}$),
normalized to the point-like QED cross section,
compared to the standard model prediction $R^{SM}_{ll}$.
The dotted and dashed lines correspond to the
one standard deviation experimental errors for
$R_{\mu\mu}$ and $R_{\tau\tau}$ respectively.
The shaded areas, enclosed between the solid lines
($M_\beta=200\>$GeV) and the dot-dashed lines
($M_\beta=300\>$GeV)
depict the predictions for lepton pair
production at the TRISTAN c.m. energy $\sqrt{s}=58\>$GeV
in the unconventional assignments $\E$ models.
The results are given
for a general $Z_\beta$ from $\E$, as a function of $\sin\beta$.
$R_{16\to 16}$ refers to the case in which
the L-components of both the initial and final leptons
are assigned to the {\bf 16} representation of $SO(10)$,
and similarly for $R_{16\to 10}$ and $R_{10\to 10}$.
}
}
$$
}
\vskip -.6truecm
\endinsert
\interlinea
\noindent
In Fig. 1 we compare the theoretical values for
$R_{16\to 16}$,
$R_{16\to 10}$ and
$R_{10\to 10}$ at $\sqrt{s}=58\,$ GeV c.m. energy,
with the SM prediction
$R^{SM}_{ll}=1.053$  (heavy solid line),
and with the TRISTAN experimental data
$R_{\tau\tau}=1.026\pm 0.037$ and
$R_{\mu\mu}=0.982\pm 0.036$.
These figures have been obtained by combining the
results of the AMY, TOPAZ and VENUS
collaborations given in Ref. [\cite{kek}].
In deriving the averages we have assigned a common systematic error
of $\pm 0.030$ for the uncertainty in the luminosity.
Fig. 1 shows that the measured values of $R_{\tau\tau}$
and $R_{\mu\mu}$ are both lower than the SM prediction.
However, while the value of
$R_{\tau\tau}$ is consistent with the SM
within one standard deviation,
$R_{\mu\mu}$ is about two standard deviations off the
expected value.
The shaded areas show the predictions for the three ratios
$R_{16\to 16}$,
$R_{16\to 10}$ and
$R_{10\to 10}$
for a $Z_\beta$ mass ranging between 200 GeV and 300 GeV.
This range coincides with the range of the lower bounds
on $M_\beta$ obtained in the framework of conventional E$_6$ models.
For example, for the models usually referred to as $\psi$,
$\eta$ and $\chi$ models which correspond
to the particular values $\sin\beta=-\sqrt{5/8}$, 0, 1
[\cite{rizzo-e6}],
the most conservative direct bounds
are respectively 200, 230 and 280 GeV.
These bounds have been obtained
at hadron colliders
from the limits on the
process $p\bar p\to Z_\beta\to l^+l^-$,
by assuming $Z_\beta$ decay to all allowed
fermions and supersymmetric fermions [\cite{zp-direct}].
Other indirect limits have been obtained from
high precision electroweak data by
analysing the
$Z_\beta$ indirect effects on NC observables.
The indirect bounds also  suggest $M_\beta \gtrsim 200\>$GeV
for all the values of the parameter $\beta$ [\cite{zp-new}].
Clearly the limits derived from analyses based on the conventional
scheme cannot be straightforwardly applied to the
$Z_\beta$ of UA models,
since in the present case a large number of fermion couplings
could be different.
However, we have no reason to expect that by assuming UA
the bounds could be greatly strengthened or relaxed, and
hence we will take the quoted limits as a reasonable guess for
the lower bounds on $M_\beta$ also in the UA schemes.

{}From Fig. 1, it is apparent that
in the presence of a light $Z_\beta$,
the experimental data
on $R_{\mu\mu}$ would be better accounted for by either
assigning both ``$e_L$" and ``$\mu_L$" to the {\bf 16} representation of
$SO(10)$ (as in conventional $\E$ models),
and for values of $\sin \beta$ centered around zero, or by
assigning both these leptons to the {\bf 10}
and for $\sin \beta$ close to unity.
At the same time, for any choice of the assignments
and for any value of $\beta$
the various $R$ are in good agreement
with the
experimental value of $R_{\tau\tau}$. Only a small region
in the vicinity of $\sin \beta \sim 0.6$
is slightly disfavoured if the
assignments
``$e_L$"$\in$  {\bf 16} and  ``$\tau_L$"$\in$  {\bf 10}
are chosen.

\midinsert
{
\vskip 1truecm
\epsfxsize=14.0truecm 
\epsffile{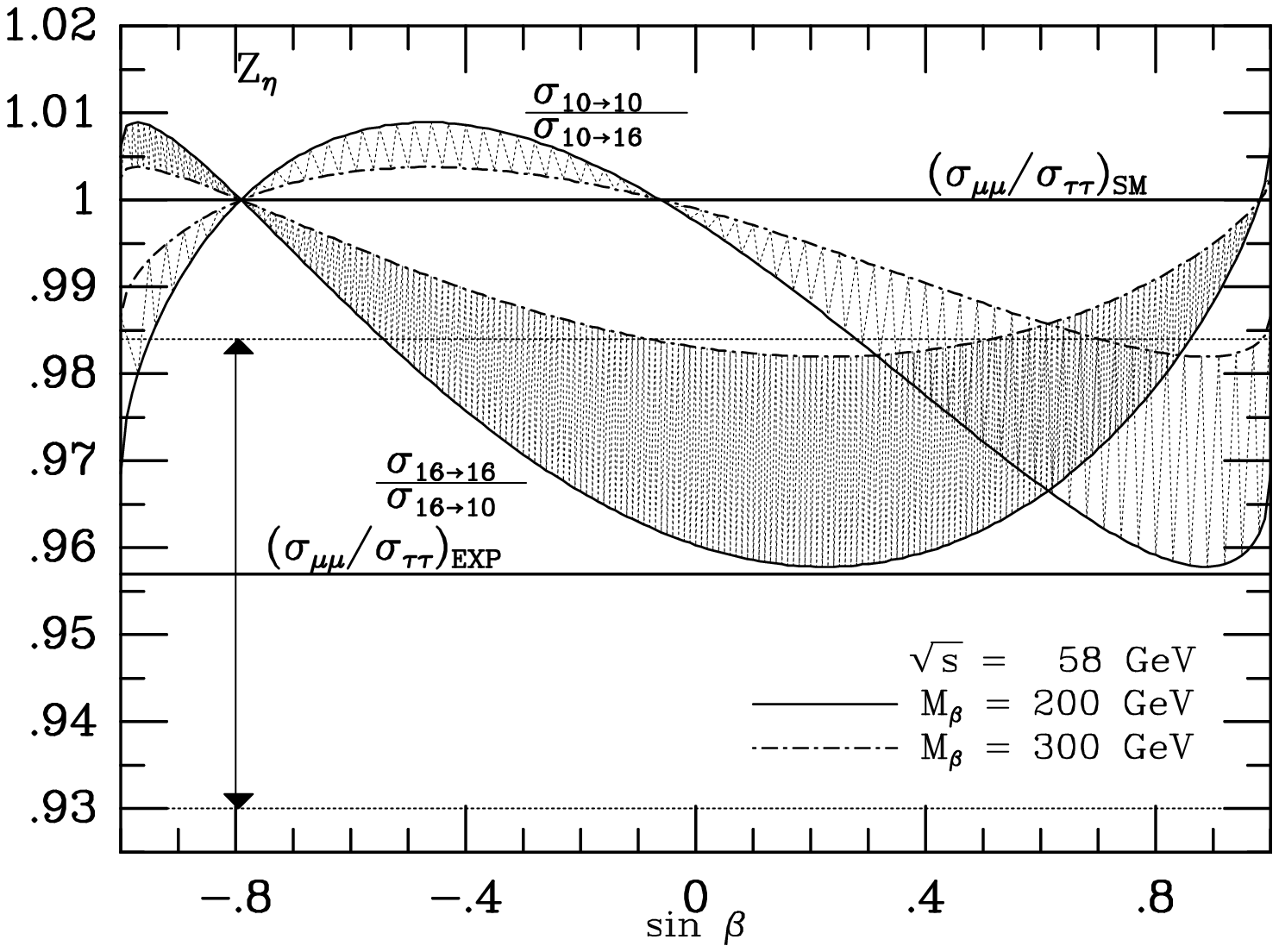}
\vskip -10.5truecm
\baselineskip=14pt
$$
\vbox{
\hsize= 13.3truecm
\noindent
FIG. 2.
{\nine
The TRISTAN result for the ratio of
$\mu$ to $\tau$ pair productions
$(\sigma_{\mu\mu}/\sigma_{\tau\tau})_{\rm EXP}$
compared to the standard model prediction
$(\sigma_{\mu\mu}/\sigma_{\tau\tau})_{\rm SM}=1$.
The dotted lines correspond to the
one standard deviation experimental error.
The shaded areas, enclosed between the solid lines
($M_\beta=200\>$GeV) and the dot-dashed lines
($M_\beta=300\>$GeV)
depict the predictions for the ratio of pair
productions of two different lepton flavors
at $\sqrt{s}=58\>$GeV,
in the unconventional assignments $\E$ models.
The results are given
for a general $Z_\beta$ from $\E$, as a function of $\sin\beta$.
$\sigma_{16\to 16}$ refers to the case in which
the L-components of both the initial and final leptons
are assigned to the {\bf 16} representation of $SO(10)$,
and similarly for $\sigma_{16\to 10} = \sigma_{10\to 16}$
and $\sigma_{10\to 10}$.
}
}
$$
}
\vskip -.6truecm
\endinsert
\interlinea
\noindent
One of the most spectacular signals of UA models
would be a deviation of the ratio $\rho_{\mu/\tau}\equiv
\sigma (e^+e^- \to \mu^+\mu^-)/ \sigma (e^+e^- \to \tau^+\tau^-)$
from unity.
Since many systematic errors, take for example
the uncertainties in the luminosity measurements,  cancel
in this ratio, the experimental error
is statistically dominated, implying a
very transparent significance for such a measurement.
At the $Z_0$ resonance,
the measured value for this observable
 $\rho_{\mu/\tau}=0.998\pm 0.006$ [\cite{lep-93}],
in striking agreement with $\mu$--$\tau$ universality.
Undoubtedly it would be difficult to accommodate
any large deviation from unity at $\sqrt{s}\ne M_Z$
by means of some mechanism different from the one discussed here.
For this reasons we believe that if a value
$\rho_{\mu/\tau}\ne 1$ is measured off $Z_0$ resonance,
this  would represent
a very clean and almost unmistakable signature of the UA
models.

According to the assignments in \req{3.5}, and without referring to
any specific lepton flavor,
we can write two expressions for
this ratio  which deviate from unity:
$\rho_{16} \equiv {\sigma_{16\to 16}/ \sigma_{16\to 10}}$
and
$\rho_{10}\equiv {\sigma_{10\to 10}/ \sigma_{10\to 16}}$, where the
subscripts label the assignment for the L-electron in the initial
state.
Fig. 2 shows the two ratios $\rho_{16}$
and $\rho_{10}$ compared to the combined TRISTAN
measurement $\rho_{\mu/\tau}=0.957\pm 0.027$.
This value is about 1.6 standard deviation off the value of unity
predicted by any model which assumes lepton universality.
Again it is apparent that the experimental data
can be better accounted for by taking
$$
\eqlabel{3.6}
\eqlabel{3.7}
\eqalignno{
``e_L",``\mu_L" &\in{\bf 16}, \quad ``\tau_L"\in  {\bf 10}
\qquad {\rm and} \qquad -0.5\lesssim\sin\beta\lesssim 0.8 &\req{3.6} \cr
{\rm or} \phantom{``\mu_L" }& & \cr
``e_L",``\mu_L" &\in{\bf 10}, \quad ``\tau_L"\in  {\bf 16}
\qquad {\rm and} \qquad \sin\beta\gtrsim0.4\, &\req{3.7} \cr
}
$$
and assuming $M_\beta \lesssim 300$ GeV.
We note that the  set of assignments in \req{3.6}
coincides with the assignments in the model
discussed in Ref. [\cite{f}].

For the particular value $\sin\beta=-\sqrt{5\o 8}$,
which corresponds to the rank 5 $\eta$ model
[\cite{rizzo-e6}],
the cross section \req{3.1} is invariant with respect to
the different choices of the assignments.
This is apparent from Fig. 1, and in particular
Fig. 2 shows that in this case lepton universality is preserved.
This follows from the fact
that for this value of $\beta$,
the $Q_\eta$ charges for the
the ${\bf \bar 5}_{({\bf 10})}$ and for
the ${\bf \bar 5}_{({\bf 16})}$ are equal [\cite{rizzo-e6},\cite{lfc}]
(this is true also for
 $Q_\eta({\bf 1}_{({\bf 1})})$ and $Q_\eta({\bf 1}_{({\bf 16})})$)
implying that for all the leptons
the couplings to the $Z_\eta$
are the same independently of the UA.

\midinsert
{
\vskip 1truecm
\epsfxsize=14.0truecm 
\epsffile{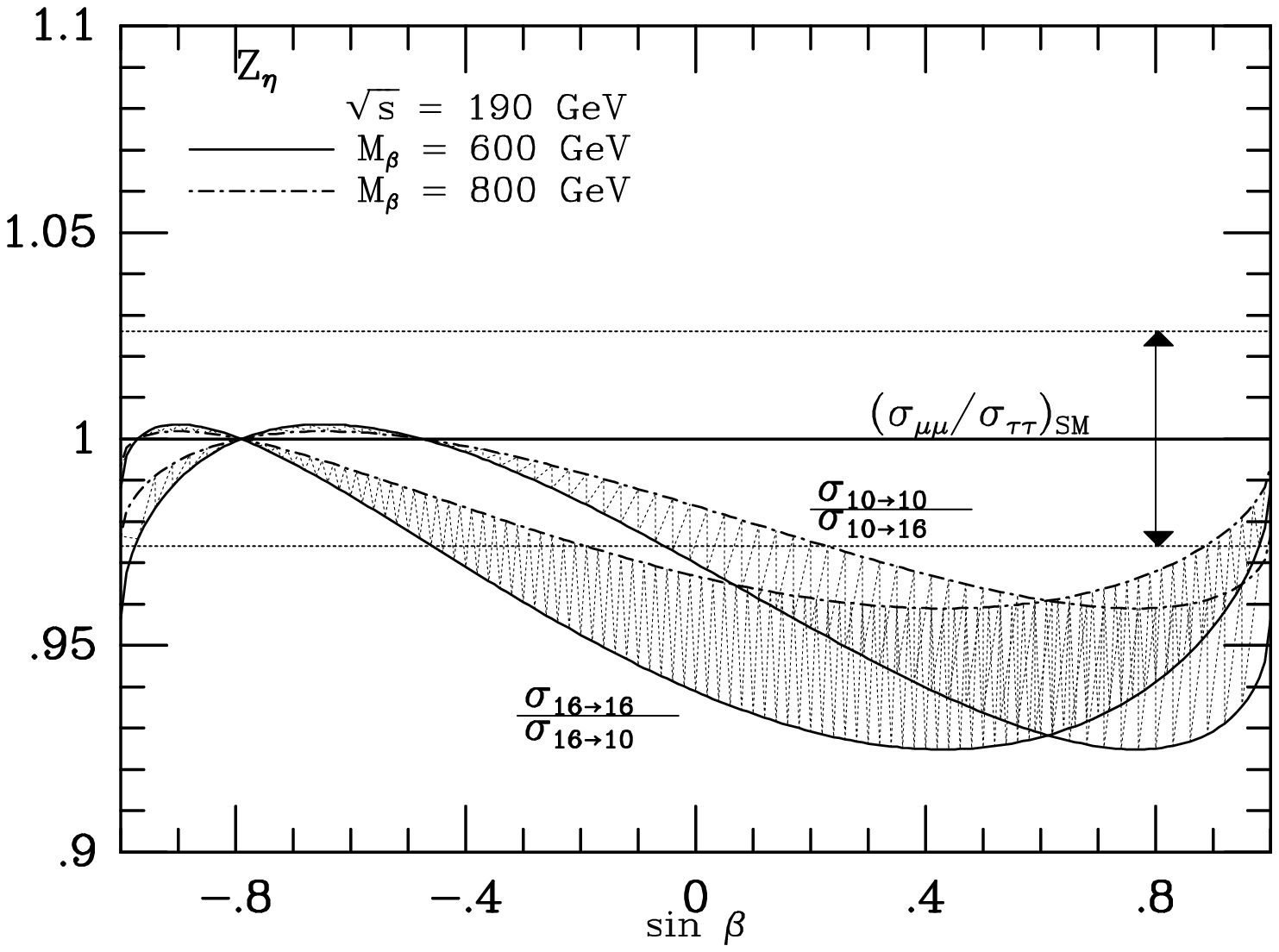}
\vskip -10.5truecm
\baselineskip=14pt
$$
\vbox{
\hsize= 13.3truecm
\noindent
FIG. 3.
{\nine
The predictions for the ratio of pair
productions of two different lepton flavors
in the unconventional assignments $\E$ models at the
LEP-2 c.m. energy $\sqrt{s}=190\>$GeV  (shaded areas),
compared to the standard model prediction (thick solid line).
The dotted lines depict  the expected
one standard deviation experimental error, based
on an integrated luminosity of 500 pb$^{-1}$.
The solid lines enclosing the shaded areas correspond to
$M_\beta=600\>$GeV while the dot-dashed lines correspond to
$M_\beta=800\>$GeV. The results are given for a general $Z_\beta$ from
$\E$, as a function of $\sin\beta$.
}
}
$$
}
\vskip -.6truecm
\endinsert
\interlinea
\noindent
In Fig. 3 we depict the theoretical values of
$\rho_{16}$  and $\rho_{10}$
for $M_\beta$ in the range 600 -- 800 GeV,
and for $\sqrt{s}=190\>$GeV, corresponding to
the c.m. energy at LEP-2.
The dotted lines depict the one standard deviation statistical error
achievable with 500 pb$^{-1}$ of integrated luminosity corresponding to
one year run [\cite{lep2-perkins}]
($\sim 3\times 10^3$ leptonic events per flavor).
It is apparent that the signature of UA models could be
easily recognized for $Z_\beta$ bosons corresponding to
most of the $\sin\beta$  positive values.

\midinsert
{
\vskip 1truecm
\epsfxsize=14.0truecm 
\epsffile{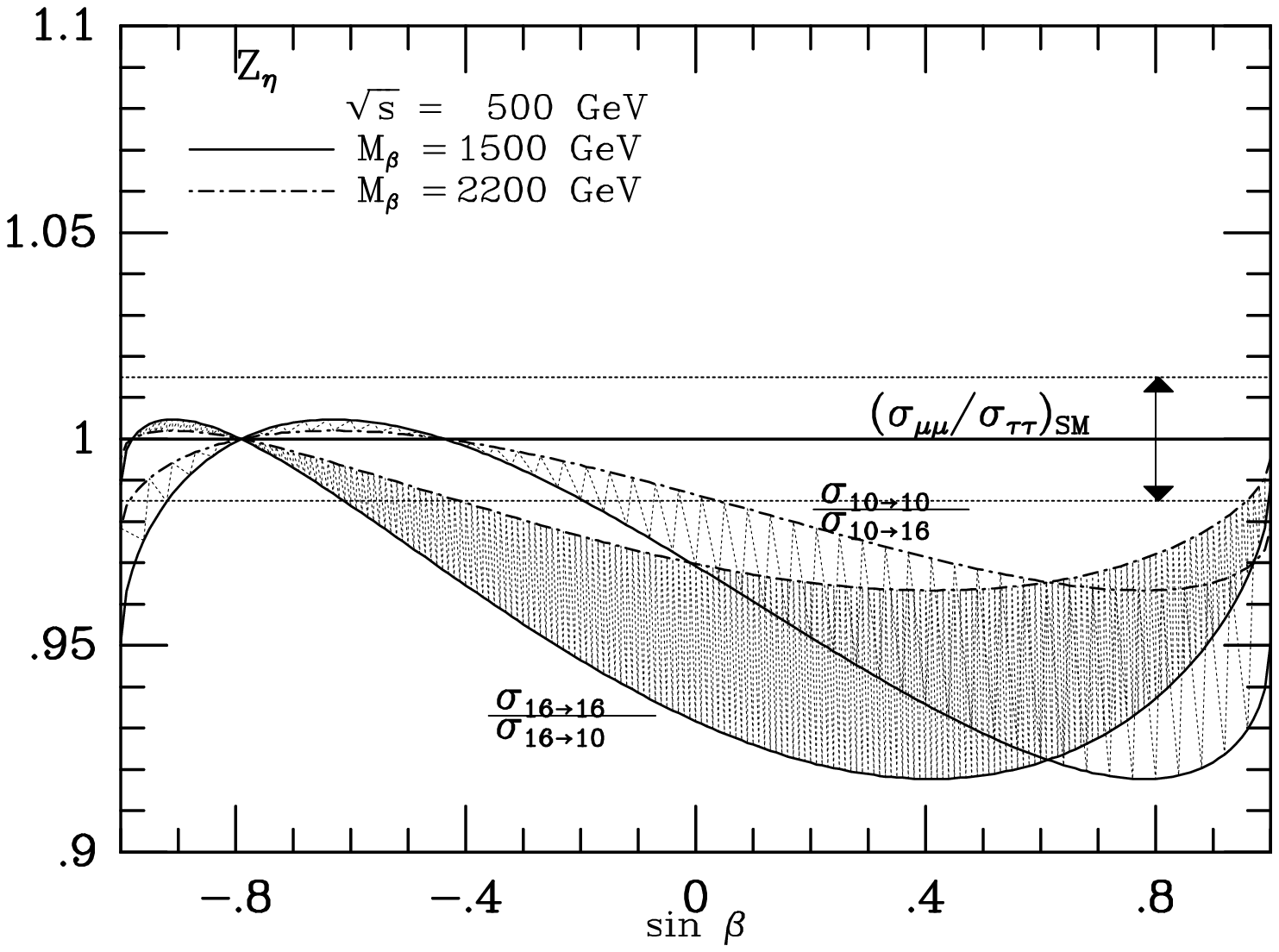}
\vskip -10.5truecm
\baselineskip=14pt
$$
\vbox{
\hsize= 13.3truecm
\noindent
FIG. 4.
{\nine
Same as Fig. 3, at the NLC c.m. energy $\sqrt{s}=500\>$GeV, for
$M_\beta=1500\>$GeV (solid lines) and $M_\beta=2200\>$GeV  (dot-dashed
lines). The dotted lines depict  the expected one standard deviation
experimental error, based on an integrated luminosity of 20 fb$^{-1}$
and including an efficiency cut.
}
}
$$
}
\vskip -.6truecm
\endinsert
\interlinea
\noindent
Fig. 4 depicts the ratios $\rho_{16}$  and $\rho_{10}$
for $M_\beta$ in the range 1500 -- 2200 GeV,
and for $\sqrt{s}=500\>$GeV, corresponding to
the NLC c.m. energy.
The one standard deviation error
corresponds to a statistics of
$\sim 8.6\times 10^3$
leptonic events per flavor,
based on an integrated luminosity of 20 fb$^{-1}$
(one year run) and taking into account the efficiency for
a cut to suppress the two-photon background
[\cite{nlc-djouadi}].
The violation of lepton universality, intrinsic of the
UA models, would produce striking effects for
a $Z_\beta$ as heavy as $\sim 2\>$TeV and for
values of $\sin\beta$ not too close to the $\eta$ model.

We have seen that by measuring
the various quantities $R_{ll}$
and $\rho_{l/l^\pr}$, it will be easy to
detect the effects of the UA.
However, from Figs. 1--4 it is also apparent that
these observables alone would not be sufficient to
determine the exact pattern of lepton assignments, since
different assignments can account for the same set of experimental data
by means of a different choice of the $\beta$ parameter.
Working out a procedure for identifying univocally the correct
pattern of assignments is beyond the scope of the present analysis,
however,
we believe that once signals of UA are detected in the leptonic cross
sections, a measurement of the various leptonic asymmetries
would be quite effective to achieve this result.
Also, we would like to point out that if some mechanism resulting in
generation dependent assignments for the lepton doublets
is effective, quite naturally the same mechanism would imply UA for the
$d$-type $SU(2)$ singlet quarks as well. For example
in the model discussed in Ref. [\cite{f}]
the assignments
$``e_L",``\mu_L" \in{\bf 16}$ and $``\tau_L"\in  {\bf 10}$
did imply, for the consistency of the model,  the UA
$``d^c_L",``s^c_L" \in{\bf 10}$ and $``b^c_L"\in  {\bf 16}$
in the quark sector.
Clearly, due to the experimental difficulties in tagging the quark
flavors,
identifying UA for the $d$-quarks would be a much harder task.

\chap{Conc} {IV. CONCLUSIONS}{}

In conclusion we have examined the possibility of
detecting with the present and future $e^+e^-$
colliders,
signals of models predicting UA for the charged leptons.
We have shown that a class of
models based on the gauge group $\E$, in which the known
$SU(2)$ lepton doublets
are embedded
in the fundamental representation of the group
in a generation dependent way,
would result in a unique
type of violation of lepton universality,
which is induced by the exchange
of new $Z_\beta$ bosons.
In agreement with LEP-1 data, no
observable effects are predicted at the $Z_0$ resonance,
however, some signals could be detected off $Z_0$ resonance.
For example, we have shown that a few anomalies
in the production rate of leptons, as well a
hint of violation of $\mu$--$\tau$ universality which have been
observed at the TRISTAN collider,
could be well accounted for in the UA scenarios.
As we have discussed, though these anomalies are not statistically
compelling, it will not be easy to find an alternative
particle physics mechanism that could simultaneously account for
the LEP-1 and the TRISTAN observations.
However, the mechanism proposed here would be effective
only if the $Z_\beta$ mass is not much heavier than $\sim 300\>$GeV.
Though this value is still consistent with the direct limits from
the $p\bar p$ collider [\cite{zp-direct}],
in the near future the data obtainable
at Tevatron will be able to probe or rule out such an explanation
of the leptonic cross section anomalies
[\cite{zp-direct-new}].

We have also discussed the discovery potential for this class of
models at LEP-2, operating at 190 GeV c.m. energy, and at the NLC,
operating at 500 GeV c.m. energy.
We have shown that at these future colliders,
striking effects of lepton universality violation
resulting from the various UA, could be observed
up to $M_\beta \sim 800$ GeV (LEP-2) and
$M_\beta \sim 2200$ GeV (NLC)
for most of the values of the model dependent parameter $\beta$.

\chap{Ackn} {ACKNOWLEDGEMENTS}{}

It is a pleasure to thank the Theory Group at KEK and the Particle
Physics Group at the Yukawa Institute in Kyoto for their kind
hospitality during the initial stage of this work.
I also thank K. Hagiwara and G. Kane for some useful discussions,
and J. F. Dodge, I. Rothstein, and D. Tommasini,  for a
critical reading of the manuscript.

\vfil\eject

\null
\centerline{\title References}
\baselineskip 8pt

\vskip .8truecm

\biblitem{f}
E. Nardi, Report No. UM-TH-93-09 to be published on \prd{48} n.7
(1 October 1993). \par

\biblitem{see-saw}
M. Gell-Mann, P. Ramond, and R. Slansky, in {\it
Supergravity}, Proceedings of the Workshop, Stony Brook, New
York,1979, edited by
F. van Nieuwenhuizen and D. Freedman, (North
Holland, Amsterdam, 1979) p.~315; \hbup
T. Yanagida, {\it Proceedings of the
Workshop on Unified Theory and Baryon Number of the Universe},
Tsukuba, Japan, 1979, edited by A. Sawada and A. Sugamoto,
(KEK Report No. 79-18, Tsukuba, 1979). \par

\biblitem{rizzo-e6}
See J.L. Hewett and T.G. Rizzo, \prep{183} 195 (1989) and references
therein. \par

\biblitem{zp-kek}
K. Hagiwara, R. Najima, M. Sakuda and T. Terunuma; \prd{41} (1990)
815. \par

\biblitem{nandi-sarkar}
S. Nandi and U. Sarkar, \prl{56} (1986) 564;\hbup
M. Cveti\v c and P. Langacker, \prd{46} (1992) R2759.\par

\biblitem{ellis-e6}
B.A. Campbell \ea, \ijmpa{2} 831 (1987). \par

\biblitem{MNS}
A. Masiero, D.V. Nanopoulos and A.I. Sanda,
\prl{57} 663 (1986).  \par

\biblitem{branco}
G.C. Branco and C.Q. Geng, \prl{58} 969 (1986).\par

\biblitem{witten-yuk}
E. Witten, \npb{258} 75 (1985). \par

\biblitem{MSW}
L.~Wolfenstein, \prd{17} 2369 {1978}; {\bf D20}, 2634 (1979); \hbup
S.~P.~Mikheyev and A.~Yu.~Smirnov,
{\it Yad.  Fiz.} {\bf 42}, 1441 (1985); \ncim{{9C}} 17 (1986).\par

\biblitem{e6-super}
D. Gross, J. Harvey, E. Martinec and R. Rhom, \prl{54} 502 (1985);
\npb{256} 253 (1985); \ib {\bf B 267} (1986) 75; \hbup
P. Candelas, G. Horowitz, A. Strominger and E. Witten, \npb{258}
46 (1985); E. Witten, \plb{149} 351 (1984). \par

\biblitem{weinberg-fc}
S. Glashow and S. Weinberg, \prd{15} 1958 (1977).\par

\biblitem{ll1}
P. Langacker and D. London, \prd{38} 886 (1988). \par

\biblitem{lep-93}
See A. Gurtu, invited talk given at the
{\it 10th DAE Symposium on High Enenrgy Physics},
Bombay, India, 1992, Report No. TIFR/EHEP 93-1 (unpublished). \par

\biblitem{lfc}
E. Nardi, \prd{48} 1240 (1993). \par

\biblitem{kek}
AMY Collaboration, T. Kumita;
TOPAZ Collaboration, B.L. Howell;
VENUS Collaboration, T. Sumiyoshi;
in {\it Proceedings of the Workshop on TRISTAN Physics at High
Luminosities}, KEK, Tsukuba, Japan, 1992,
edited by M. Yamauchi (KEK Proceedings No. 93-2);  \hbup
see also A. Maki, in {\it Colmar 1991, Proceedings of the
11th International Conference on Physics in Collision}
Colmar, France, 1991 (KEK Report No. 91-100). \par

\biblitem{lep2-perkins}
D. H. Perkins, in {\it Proceedings of the ECFA Workshop on LEP 200},
Aachen, Germany, 1986, edited by A. B\"ohm and W. Hoogland
(CERN Report No. 87--08, p.1). \par

\biblitem{zp-direct-new}
F. del Aguila, M. Cveti\v c and P. Langacker,
in {\it Proceedings of the Workshop on Physics and Experiments with
Linear $e^+e^-$ colliders},
Waikoloa, HI, 1993, edited by F. Harris,
Report No. UPR-0583-T (Unpublished). \par

\biblitem{nlc-djouadi}
A. Djouadi \ea,
in {\it Proceedings of the Workshop on $e^+e^-$ Collisions at
500 GeV}, Munich, Annecy, Hamburg, 1991,
edited by P.M. Zerwas (DESY Report No. 92-123B, p. 491). \par

\biblitem{domain}
Ya. B. Zeldovich, I. YU. Kobzarev and L. B. Okun, \plb{50} 340
(1974).\par

\biblitem{ewinflation}
L. Knox and M. S. Turner, \prl{70} 371 (1992). \par

\biblitem{nucleo-nu}
T. Walker, in  Texas/PASCOS 92:
{\it Relativistic Astrophysics and Particle Cosmology, Proceedings of
the 16th Texas Symposium on Relativistic Astrophysics and 3rd
Particles, Strings, and Cosmology Symposium}, Berkeley, CA, 1992,
edited by C.W. Akerlof and M.A. Srednicki (The New York Academy of
Sciences, vol. 688, New York, 1993), p. 745.  \par

\biblitem{nucleo-e6}
M.C. Gonzales-Garcia and J.W. Valle, \plb{240} 163 (1990); \hbup
J. L Lopez and D.V. Nanopoulos, \ib {\bf 241} 392 (1990); \hbup
A.E. Faraggi and D.V. Nanopoulos, Mod. Phys. Lett. {\bf A6} 61 (1991). \par

\biblitem{cecilia}
C. Jarlskog, Nucl. Phys. {\bf A518}, 129 (1990). \par

\biblitem{zp-direct}
CDF Collaboration, F. Abe \ea, \prl{67} 2609 (1991); \ib {\bf 68},
1463 (1992). \par

\biblitem{zp-new}
E. Nardi, E. Roulet and D. Tommasini, \prd{46} 3040 (1992); \hbup
P. Langacker and M. Luo, \ib {\bf 45} 278 (1992); \hbup
F. del Aguila, W. Hollik, J.M. Moreno and M. Quir\'os, \npb{372}
 3 (1992); \hbup
J. Layssac, F.M. Renard and C. Verzegnassi, \zpc{53} 97 (1992); \hbup
M.C. Gonzalez Garc\'\i a and J.W.F. Valle; \plb{259} 365 (1991); \hbup
G. Altarelli \ea, \ib {\bf 263} 459 (1991). \par

\biblitem{atmnu-matter}
E. Akhmedov, P. Lipari and M. Lusignoli; University ``La Sapienza" -
Rome Report No. 912 (November 1992). \par

\biblitem{KAM92a}
K.S. Hirata  {\it et al.} (Kamiokande-II Collaboration),
\plb{280} 146 (1992). \par

\biblitem{IMB92a}
R. Becker-Szendy {\it et al.} (IMB Collaboration), \prd{46} 3720
(1992); see also D. Casper {\it et al.}, \prl{66} 2561 (1991). \par

\biblitem{KAM92c} E.W. Beier  \ea \ \plb{283} 446 (1992). \par

\biblitem{IMB92b} R. Becker-Szendy {\it et al.} (IMB Collaboration),
\prl{69} 1010 (1992). \par

\biblitem{Baksan91}
Baksan Collaboration,
M.M. Boliev {\it et al.}
in {\it Proceedings of the 3rd International Workshop on
Neutrino Telescopes,} Venice, Italy, 1991, edited by Milla Baldo Ceolin
(Istituto Nazionale di Fisica Nucleare, Padova, 1991), p. 235 \par

\biblitem{leprad}
M. Consoli and W. Hollik, in {\it Z physics at LEP 1} Vol. 1,
edited by G. Altarelli \ea, CERN Report No. 89--08, p. 7;
G. Burgers and F. Jegerlehner, \ib, p. 55; \hbup
G. Burgers and W. Hollik, in {\it Polarization at LEP} Vol. 1,
edited by G. Alexander \ea, CERN Report No. 88--06, p. 136; \hbup
D.C. Kennedy and B.W. Lynn, \npb{322} (1989) 1. \par

\insertbibliografia
\vfill\eject
\bye